\documentclass[aps,prl,reprint,groupedaddress]{revtex4-2}
\usepackage{graphicx}
\usepackage{amsmath}
\usepackage[colorlinks,linkcolor=blue,citecolor=blue,urlcolor=black]{hyperref}

\begin{document}


\title{Higher-order network adaptivity: co-evolution of higher-order structure and spreading dynamics}


\author{Longzhao Liu$^{1,3,5,6,7}$}
\author{Hongwei Zheng$^{8}$}
\author{Zhihao Han$^{1,3}$}
\author{Xin Wang$^{1,3,5,6,7}$}
\email{wangxin\_1993@buaa.edu.cn}
\author{Shaoting Tang$^{1,2,3,4,5,6,7}$}
\email{tangshaoting@buaa.edu.cn}
\affiliation{$^1$Institute of Artificial Intelligence, Beihang University, Beijing 100191, China}
\affiliation{$^2$Hangzhou International Innovation Institute, Beihang University, Hangzhou 311115, China}
\affiliation{$^3$Key laboratory of Mathematics, Informatics and Behavioral Semantics, Beihang University, Beijing 100191, China}
\affiliation{$^4$Institute of Medical Artificial Intelligence, Binzhou Medical University, Yantai 264003, China}
\affiliation{$^5$Zhongguancun Laboratory, Beijing 100094, China}
\affiliation{$^6$Beijing Advanced Innovation Center for Future Blockchain and Privacy Computing, Beihang University, Beijing 100191, China}
\affiliation{$^7$State Key Laboratory of Complex \& Critical Software Environment, Beihang University, Beijing 100191, China}
\affiliation{$^8$Beijing Academy of Blockchain and Edge Computing, Beijing 100085, China}


\begin{abstract}
The co-evolution of structure and dynamics, known as adaptivity, is a fundamental property in various systems and drives diverse emergent behaviors. However, the adaptivity in previous works is primarily stemmed from pairwise situations, while is insufficient to capture ubiquitous higher-order characteristics of real systems. Here, we introduce higher-order network adaptivity to model the co-evolution of higher-order structure and spreading dynamics, and theoretically analyze the thresholds and spreading sizes. Results demonstrate that both pairwise-like and higher-order adaptivity consistently increase spreading thresholds, but surprisingly produce completely opposing qualitative effects. Specifically, contrary to pairwise-like adaptivity, higher-order adaptivity not only reduces or even eliminates the bistable region, but also leads to shifts of phase transitions from discontinuous to continuous. These findings are validated on both synthetic and real hypergraphs. Our work introduces an idea of higher-order adaptivity and highlights its fundamental differences from pairwise-like adaptivity, advancing further researches of adaptive higher-order systems.
\end{abstract}

\maketitle

Complex systems, ranging from social to natural domains, can typically be modeled as networks and the dynamics on top of them, which jointly govern the emergent behaviors \cite{boccaletti2006complex, dorogovtsev2008critical, barrat2008dynamical, zhu2024reputation}. In paricular, networks usually co-evolve with the dynamics of nodes' states in real-world systems \cite{berner2023adaptive, gross2008adaptive, liu2020homogeneity}. For instance, in the spreading of infectious diseases, healthy individuals may avoid contacts with infected ones, causing changes of network structure, which in turn affects the spreading dynamics. The mutual influence between structure and dynamics, known as adaptivity, is identified as core mechanism in various systems and drives rich phenomena, such as bistability \cite{gross2006epidemic}, echo chamber \cite{wang2020public} and cooperation \cite{fu2008reputation}. However, these works mainly concentrate on pairwise situations which assume all interactions between units are dyadic.

Recently, higher-order interactions among multiple units have been empirically observed across diverse systems, and exhibit fundamental differences from pairwise interactions at multiple levels \cite{battiston2020networks, santoro2023higher, grilli2017higher, burgio2024triadic, lambiotte2019networks}. Structurally, such interactions require characterizations using higher-order structure such as hyperedges and simplicial complexes, which can not be decomposed into pairwise links \cite{shi2022simplicial, boccaletti2023structure}. Dynamically, higher-order interactions are typically manifested as nonlinear dynamical mechanisms that lead to entirely new emergent behaviors beyond pairwise interactions in various real-world systems \cite{ferraz2024contagion, battiston2021physics, majhi2022dynamics, malizia2025hyperedge,neuhauser2020multibody}. For instance, higher-order spreading mechanism can induce the emergence of bistability and discontinuous phase transitions \cite{iacopini2019simplicial}. Higher-order games may result in explosive transition to cooperation \cite{civilini2024explosive}. In addition, recent studies provide frameworks of temporal higher-order networks, and reveal its unique topological properties and impacts on dynamical behaviors like percolation time and contagion thresholds \cite{di2024percolation, petri2018simplicial, han2024probabilistic, iacopini2024temporal}. Despite the progress, most works study higher-order network and dynamics separately, whereas real-world systems usually exhibit co-evolution of higher-order structure and dynamics, namely higher-order adaptivity. Take the spreading of infectious diseases as a paradigmatic example \cite{st2021universal, wang2024epidemic, chen2024simplicial}. Individuals tend to leave groups with infection risks \cite{gross2006epidemic, marceau2010adaptive}, and the possibility of group breakdown may nonlinearly increases as the number of infected members grows \cite{burgio2025characteristic}. For instance, during Covid-19, individuals in low-risk groups with few infected members adopt minimal protective measures, while high-risk areas implement strict lockdown \cite{lai2020effect, perra2021non, flaxman2020estimating}. This reflects the evolution of higher-order structure induced by the number of infected individuals, which in turn affects the epidemic spreading, forming a co-evolutionary process. Nevertheless, the basic rule of higher-order adaptivity and its complex impacts on dynamical behaviors remain unclear.

In this paper, we propose a higher-order adaptive spreading model to describe the co-evolution of higher-order networks and spreading dynamics. We provide a theoretical framework that could well predict outbreak sizes and thresholds. Results on both synthetic and real hypergraphs exhibit rich and robust impacts of higher-order adaptivity. Firstly, it can increase the epidemic thresholds. Secondly, compared to pairwise-like adaptivity, higher-order adaptivity counter-intuitively induces completely opposing emergent phenomena. The former enhances or induces the bistable region, while the latter reduces or even eliminates it. Moreover, although pairwise-like adaptivity triggers discontinuous transitions, higher-order adaptivity causes that phase transitions change from discontinuous to continuous. These findings, both theoretical and experimental, highlight the unique characteristics of higher-order adaptivity in complex systems.

\begin{figure}
\includegraphics[width=0.45\textwidth]{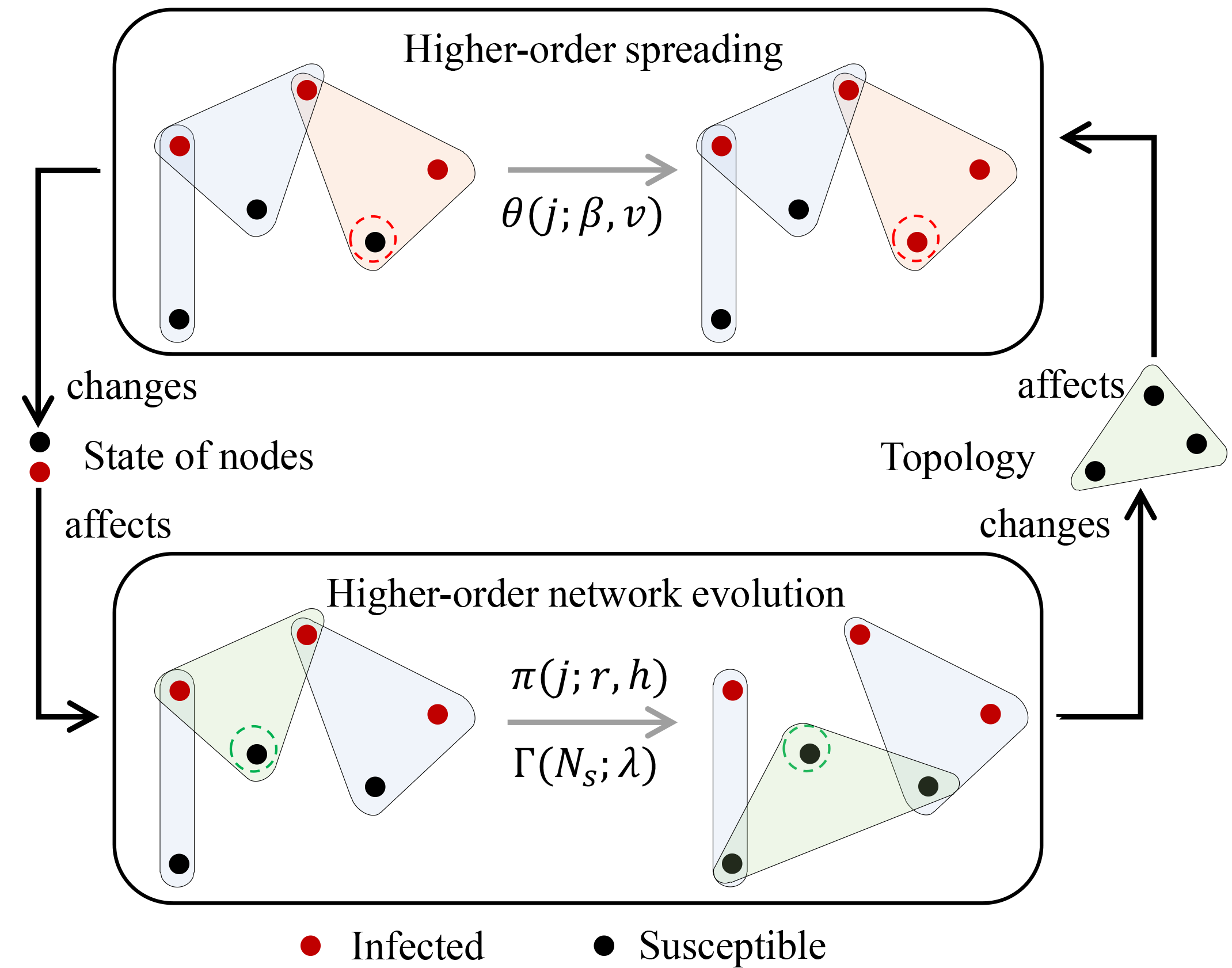}%
\caption{Higher-order adaptive spreading model. It describes the co-evolution of higher-order spreading and network topology. Specifically, the model incorporates higher-order adaptivity, characterized by power-law breaking rate of hyperedges and selected fitness, to depict the evolution of higher-order network induced by higher-order spreading dynamics. In turn, network evolution alters the topological structure, which influences the outcomes of spreading dynamics occurring on top of it. \label{model}}
\end{figure}
{\it Higher-Order Adaptive Spreading Model.} Our model is composed of two mutually interacting processes including higher-order spreading and topological evolution, as shown in Fig.\ref{model}. Firstly, we consider a typical epidemic spreading process on hypergraphs, where hyperedges of size $m$ encode higher-order interactions among $m$ individuals. Each node has two states: infected ($I$) and susceptible ($S$). Infected individuals recover at rate $\mu$, while transmitting the disease to susceptible ones at a certain infection rate. Noteworthy, recent works demonstrate the ubiquity of higher-order spreading mechanisms, which manifest as the nonlinear relationships between infection rate and the exposure to infected contacts \cite{st2021universal, st2022influential, ferraz2024contagion}. Here, we adopt the power-law infection rate
\begin{equation}
\theta(j;\beta,v)=\beta j^v
\end{equation}
where $j$ is the number of infected nodes in the hyperedge, $\beta$ is basic infection rate and $v$ adjusts nonlinearity.

Next, we introduce higher-order adaptivity to characterize how hypergraph evolves with the dynamical states of nodes. In real-world scenarios of disease transmission, people tend to adopt risk-avoidance behaviors, such as avoiding contacts with infected individuals and avoiding participating in groups with infection risks \cite{gross2006epidemic, pastor2015epidemic}. Institutions, such as school and workplace, would take lockdown measures to avoid high-risk group interactions, as was the case during the Covid-19 pandemic \cite{lai2020effect, perra2021non, flaxman2020estimating}. These behaviors and measures can be modeled as the decomposition of groups containing infected members, i.e., hyperedge breaking. Empirical observations suggest that the likelihood of taking actions may nonlinearly rise as the number of infections increases. Therefore, we assume that a hyperedge containing $j$ infected nodes and at least 1 susceptible node breaks with a power-law rate
\begin{equation}
\pi(j;r,h)=r j^h
\end{equation}
where $r$ is the basic breaking rate and $h$ represents the degree of dependence on the infection level within groups. $h=0$ means that the breaking rate is independent of the number of infected nodes, reducing to pairwise-like adaptivity. $h\neq0$ reflects more realistic higher-order properties that does not appear in pairwise situations.

Moreover, to prevent network fragmentation and for the sake of analytical tractability, we assume that a new $m-$hyperedge will form when a $m-$hyperedge breaks. Specifically, a susceptible node of the broken hyperedge will choose $m-1$ nodes to form a new one. Inspired by the fitness in evolutionary game theory \cite{nowak2006evolutionary}, we can define the probability that a susceptible node is chosen, which is
\begin{equation}
\Gamma(N_s;\lambda) = \frac{\lambda N_s}{\lambda N_s + (N-N_s)}
\label{gamma}
\end{equation}
where $N$ and $N_s$ represent the number of nodes and susceptible nodes, respectively. $\lambda\in[1,\infty)$ reflects the fitness of $S$ nodes, which is influenced by the completeness and accuracy of information about the infection status of populations. $\lambda=1$ means that individuals know no information, and thus selection is completely random, while $\lambda\rightarrow \infty$ indicates that individuals can accurately find susceptible ones at any time.

\begin{figure*}
\includegraphics[width=0.86\textwidth]{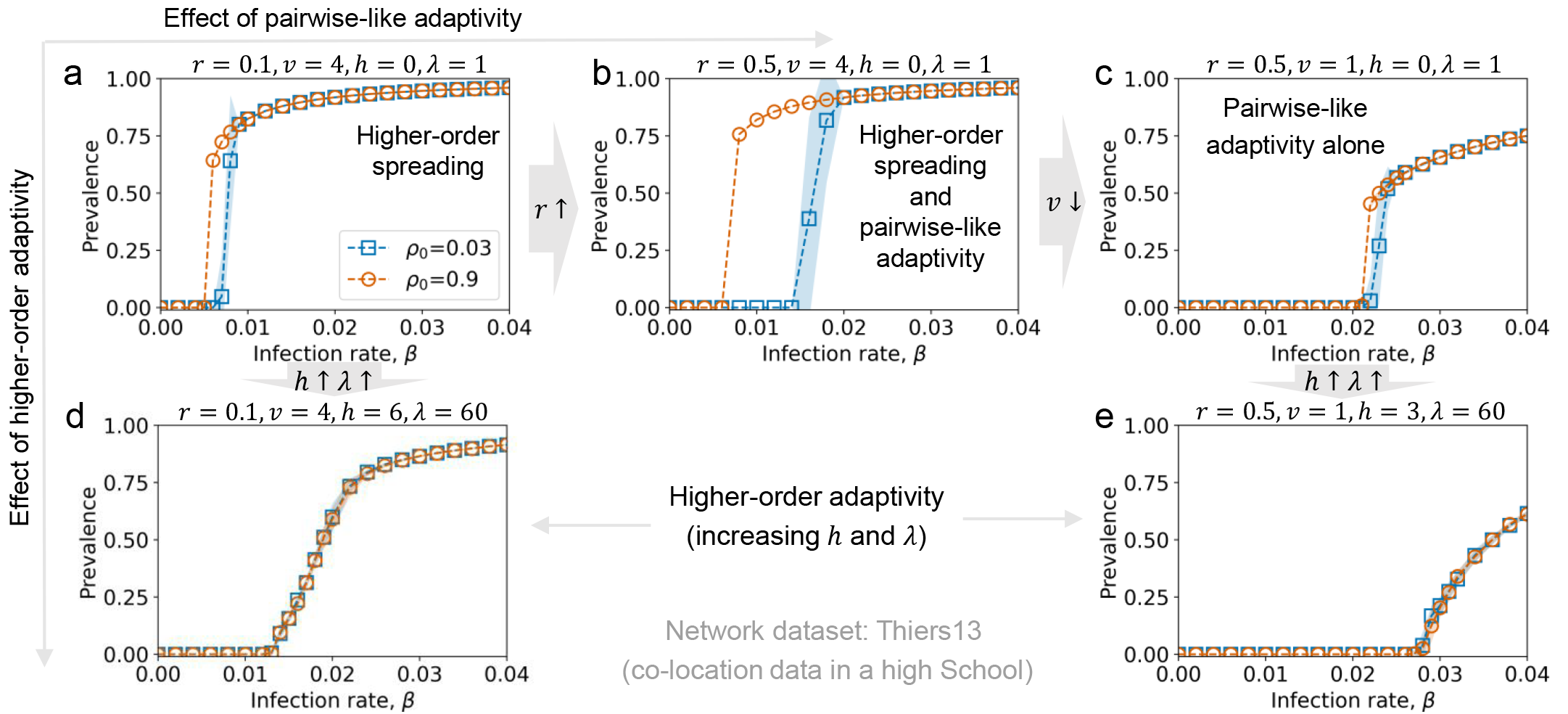}%
\caption{Results based on a real hypergraph derived from Thiers13 dataset \cite{mastrandrea2015contact}. Shown is epidemic prevalence against basic infection rate $\beta$ under different parameters. Each point is averaged over 30 runs, and the shading area represents standard deviation. $\rho_0$ denotes the initial density of infected individuals. (a)--(c) Effects of pairwise-like adaptivity, whose strength is governed by $r$. We respectively present results under different $r$ and $v$. We find that pairwise-like adaptivity can amplify or trigger the bistable region and discontinuous transitions. (d)--(e) Effects of higher-order adaptivity, characterized by $h$ and $\lambda$. Compared to (a) and (c), (d) and (e) present cases under higher values of the two parameters. Results show that higher-order adaptivity could eliminate the bistable region and discontinuous phase transitions. Other parameter: $\mu=0.2$. \label{fig2}}
\end{figure*}

{\it Results on Real Hypergraphs.} We use three human co-location datasets to derive hypergraphs \cite{mastrandrea2015contact, genois2018can, young2017construction}, and then perform model simulations on top of them using Gillespie algorithm (see Supplementary Materials, sec.II). Fig.\ref{fig2} shows results on a real hypergraph, i.e., how epidemic prevalence varies against basic infection rate $\beta$ under different parameters. Fig.\ref{fig2}(a) reproduces the impacts of higher-order spreading mechanisms, such as inducing bistability and discontinuous phase transitions \cite{iacopini2019simplicial}. The former means co-existence of epidemic extinction and outbreak where the prevalence is determined by initial density of infected individuals, while the latter reveals the catastrophic growth of prevalence around spreading thresholds. Both are significant qualitative phenomena that receive considerable attentions. Here, our primary focus is to explore how network adaptivity affects spreading thresholds and these qualitative phenomena.

First, we explore the impact of pairwise-like adaptivity, i.e., $h=0$. In this case, the strength of adaptivity is controlled by parameter $r$. From Fig.\ref{fig2}(a) to Fig.\ref{fig2}(b), we show that pairwise-like adaptivity increases spreading thresholds and enlarges the scope of bistable region. Furthermore, Fig.\ref{fig2}(c) presents the case where higher-order spreading mechanism is removed ($v=1$). We find that bistability and discontinuous transitions can be induced solely by pairwise-like adaptivity on hypergraphs, which is similar to the impacts of adaptivity on pairwise networks \cite{gross2006epidemic}. Above all, pairwise-like adaptivity not only provides another pathway to explosive transitions and bistability, but also enhances these phenomena triggered by higher-order spreading.

Then, we focus on higher-order adaptivity, which is characterized by parameters $h$ and $\lambda$ that respectively influence hyperedge breaking rate and the formation of hyperedges. Fig.\ref{fig2}(d), through comparison with Fig.\ref{fig2}(a), unveils complex impacts of higher-order adaptivity. On the one hand, higher-order adaptivity increases spreading thresholds. On the other hand, higher-order adaptivity leads to the disappearance of the bistable region and discontinuous phase transitions triggered by higher-order spreading mechanism, which is qualitatively opposite to the impacts of pairwise-like adaptivity. Such counter-intuitive findings still hold when comparing Fig.\ref{fig2}(c) and \ref{fig2}(e) where higher-order spreading is degraded ($v=1$). The robustness of these findings is also verified on other real hypergraphs (see Supplementary Materials, Fig.S1).

{\it Theoretical Framework and Results on Synthetic Hypergraphs.} What microscopic mechanism gives rise to the dramatically opposed dynamical behaviors we observe? In this session, we further develop a theoretical framework and explore the cases on synthetic uniform hypergraphs in order to get analytical insights and mechanistic explanations. We begin with the simplest $m$-uniform hypergraph, where all hyperedges contain $m$ nodes and nodes' hyperdegrees follow a Poisson distribution. Despite the straightforward setup, there is still lack of theoretical framework for higher-order co-evolutionary dynamics due to the complexity arising from mutual influences between structure and dynamics. Here, inspired by pair approximation \cite{gross2006epidemic}, we introduce hyperedge-based approximation method to characterize the dynamics of hyperedges.

Let $N_i(t)$ denote the number of infected individuals at time $t$. $l_{m,k}(t)$ denotes the number of $m$-hyperedges containing $k$ infected nodes. Then, the average rate, that a $S$ node becomes infected, can be approximated by
\begin{equation}
\Theta = \sum_{j=1}^{m} \frac{(m-j)l_{m,j}}{N-N_i} \beta j^v
\label{eq1}
\end{equation}
where $(m-j)l_{m,j}/(N-N_i)$ represents the probability that the $S$ node belongs to $m$-hyperedges containing $j$ infected nodes. Thus, the dynamical equation for the number of infected nodes can be written as
\begin{equation}
\frac{dN_i}{dt} = -\mu N_i + (N-N_i)\Theta
\label{eq2}
\end{equation}

Eq.\eqref{eq1} suggests that $\Theta$ depends on dynamic variables $l_{m,k}(t) , 0\leq k \leq m$. Thus, analyzing system behaviors also requires the dynamical equations of $l_{m,k}(t)$. Specifically, its evolution is composed of two parts: hyperedge rewiring as well as the change of nodes' state triggered by spreading processes. Consequently, we can obtain
\begin{equation}
\begin{aligned}
\frac{dl_{m,k}}{dt} &= -(\beta k^v+\Theta) (m-k)l_{m,k} - \mu k l_{m,k} \\
&+ (\beta (k-1)^v+\Theta) (m-k+1) l_{m,k-1} \\
&+ \mu (k+1) l_{m,k+1} -r k^h l_{m,k} \\
&+ \sum_{j=1}^{m-1} l_{m,j} rj^h \binom{m-1}{k} \Gamma^{m-k-1} (1-\Gamma)^k\\
\frac{dl_{m,m}}{dt} &= -\mu m l_{m,m} + (\beta (m-1)^v + \Theta) l_{m,m-1}
\end{aligned}
\label{eq3}
\end{equation}
where $\Gamma$ varies with $N_i$ according to Eq.\eqref{gamma}. The first equation depicts the cases $k=1,2, ... , m-1$. Its first four terms represent the changes caused by spreading dynamics, while the remaining terms describe hyperedge breaking and formation. Note that the total number of hyperedges, denoted by $E$, is constant, i.e., $\sum_{j=0}^{m} l_{m,j}(t) = E$. Thus, Eqs.\eqref{eq2}--\eqref{eq3} constitute a self-contained autonomous system with $(m+1)$ dimensions. When $m=2$ and $\lambda=1$, the equations would reduce to the classical model of adaptive pairwise networks \cite{gross2006epidemic}.

Without lose of generality, we primarily concentrate on more intriguing yet still theoretically tractable higher-order cases, i.e., $m=3$. In this situation, Eqs.\eqref{eq2}--\eqref{eq3} can be expressed as a 4-dimensional system of differential equations. It has a trivial fixed point located at the origin of coordinate system. In particular, the stability condition for the fixed point corresponds to the outbreak threshold, which determines if epidemic outbreaks at small $\rho_0$. Here, we calculate its stability condition by linearizing the system around the fixed point and examining whether all eigenvalues of the Jacobian matrix have negative real parts (See Supplementary Materials, sec.IV). We then obtain the outbreak threshold, which satisfies
\begin{equation}
\beta_c = \frac{\Omega + \sqrt{\Omega^2+2^{v+3}\langle d\rangle(\mu+r)(2\mu+2^h r)}}{2^{v+2}\langle d\rangle}
\label{eq4}
\end{equation}
where
\begin{equation*}
\Omega = 2^{h+1}r(1-\langle d\rangle)-4\mu \langle d\rangle
\end{equation*}
Here, $\langle d\rangle = 3E/N$ represents the average hyperdegree in synthetic hypergraphs. Eq.\eqref{eq4} uncovers the precise relationships between the outbreak threshold and model parameters. Notably, $\beta_c$ is independent of fitness $\lambda$.

\begin{figure*}
\includegraphics[width=0.86\textwidth]{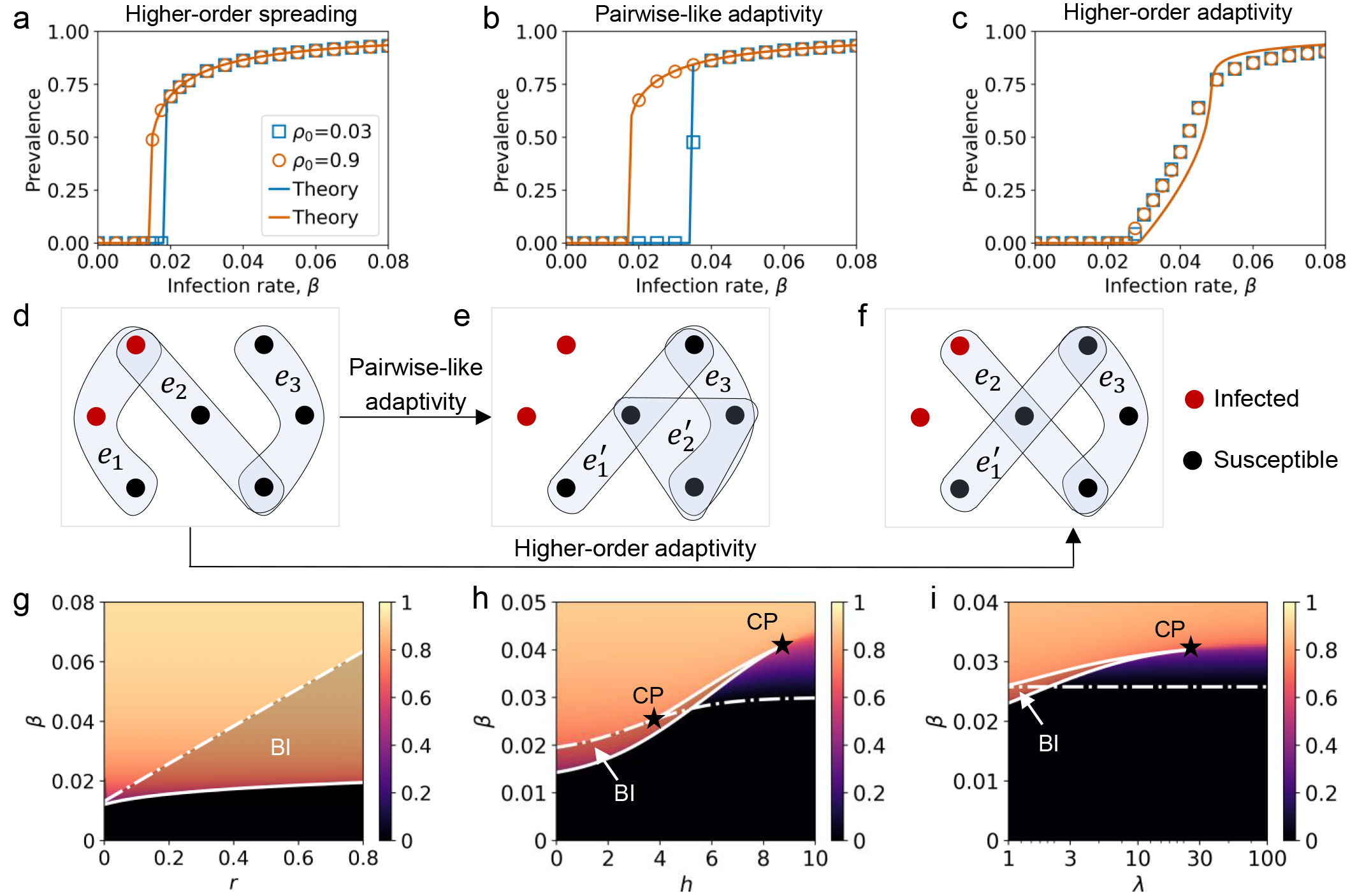}%
\caption{ Theoretical and simulation results on uniform hypergraphs. (a)--(c) Epidemic prevalence is shown under different settings: (a) $r=0.1, v=3, h=0, \lambda=1$, (b) $r=0.4, v=3, h=0, \lambda=1$, (c) $r=0.1, v=3, h=6, \lambda=60$. (d)--(f) Microscopic dynamical features of pairwise-like adaptivity and higher-order adaptivity, which could intuitively explain their rich impacts. (g)--(i) Phase diagrams with different parameters: (g) $v=3, h=0, \lambda=1$; (h) $r=0.1, v=3, \lambda=1$; (i) $r=0.1, v=3, h=4$. The dashed line, representing the outbreak threshold, is computed by Eq.\eqref{eq4} and the solid line, representing the persistence threshold or drastic-change threshold, is numerically solved by Eqs.\eqref{eq2}--\eqref{eq3}. The area they enclose is bistable region, abbreviated as BI. Moreover, there emerge cusp points (CP) in (h)--(i), which are derived through bifurcation analysis using MatCont \cite{dhooge2008new}. We find that pairwise-like adaptivity enhances bistable region, while higher-order adaptivity, characterized by parameter $h$ and $\lambda$, reduces the bistable region and alters types of phase transitions. Other parameters: $N=3000, E=6000, \mu=0.2$.  \label{fig3}}
\end{figure*}

As shown in Fig.\ref{fig3}(a)--\ref{fig3}(c), our theoretical results agree well with simulations across various parameter settings. Moreover, results on synthetic hypergraphs also show that pairwise-like adaptivity amplifies the bistable region induced by higher-order spreading, whereas higher-order adaptivity eliminates bistability and explosive transitions. We further provide intuitive explanations from microscopic perspectives in  Fig.\ref{fig3}(d)--\ref{fig3}(f). Higher-order spreading mechanism, which is the main cause of bistability in Fig.\ref{fig3}(a), operates exclusively on hyperedges containing more than one infected nodes, such as $e_1$ in Fig.\ref{fig3}(d). The introduction of pairwise-like adaptivity leads to equal breaking rate of hyperedges containing infected nodes, which produces two key effects. The first one is to disrupt structural foundation of higher-order spreading, which inhibits bistability. The second one is to induce bistability due to the reason: it triggers complete separation between $S$ and $I$ at small $\rho_0$ (see Fig.\ref{fig3}(e)), resulting in disease extinction, while at large $\rho_0$, complete separation are difficult to achieve, allowing the disease to persist. Notably, the second effect plays a dominant role. Furthermore, we explain why higher-order adaptivity produces completely opposite impacts. The most important feature of higher-order adaptivity is the nonlinear enhancement of hyperedge breaking rate with respect to the number of infected nodes. That means, hyperedges containing two infected nodes possibly break, whereas those involving only one infected node remain intact, as shown in Fig.\ref{fig3}(f). This not only disrupts the foundation of higher-order spreading but also preserves connections between $S$ and $I$ nodes. Both hinder the emergence of bistability.

To obtain more comprehensive insights, in Fig.\ref{fig3}(g)--\ref{fig3}(i), we plot phase diagrams for epidemic prevalence, with critical thresholds of the system indicated by white lines. Fig.\ref{fig3}(g) illustrates that pairwise-like adaptivity ($r$) increases spreading thresholds and the scope of bistable region. Fig.\ref{fig3}(h) systematically reveals the diverse impacts of $h$ that characterizes core feature of higher-order adaptivity. In addition to increasing spreading thresholds, we obtain two interesting phenomena. One is that the bistable region gradually shrinks and eventually disappears as $h$ increases. During the shrinkage, the nature of bistable region shifts from co-existence of extinction and outbreak to co-existence of two outbreak states. The other one is the emergence of cusp points (CP), which are boundary between different phase transitions. Specifically, as $h$ increases, both the number and the type of phase transitions alter. The system starts with a discontinuous phase transition, then followed by the coexistence of continuous and discontinuous transitions, and ultimately becomes a continuous transition (see Supplementary Materials, Fig.S2). In Fig.\ref{fig3}(i), we explore the impacts of selected fitness $\lambda$. We find that $\lambda$ does not influence the outbreak threshold, but plays an pivotal role in reducing bistable region and influencing the type of phase transitions. In addition, when excluding the influence of higher-order spreading($v=1$), higher-order adaptivity could still eliminate the bistability and discontinuous transitions caused by pairwise-like adaptivity (see Supplementary Materials, Fig.S3)

{\it Conclusions.} This work introduces higher-order adaptivity that describes the widespread co-evolution of higher-order networks and spreading dynamics in real systems. We develop a hyperedge-based theoretical framework that could well predict the system thresholds, which are validated by large-scale agent-based simulations. Of particular interest, results on both synthetic and real hypergraphs reveal the rich and counter-intuitive impacts of higher-order adaptivity. Firstly, it substantially increases spreading thresholds. Secondly, it leads to the emergence of cusp points, where the number and the type of phase transitions alter. Thirdly, from qualitative perspectives, higher-order adaptivity and pairwise-like adaptivity produce completely opposite effects: the former inhibits or eliminates bistable region, while the latter enhances or induces it. We show that such counter-intuitive phenomenon could be explained by the fundamental difference of the two adaptive mechanisms. That is, the inherent nonlinearity of higher-order adaptivity characterized by parameter $h$, where the structural evolution depends on the infection level of hyperedges.

Our findings unveil complex impacts and unique dynamical properties of higher-order adaptivity, transcending current understanding gleaned from pairwise adaptivity. While this work focuses on co-evolution of higher-order networks and spreading dynamics, the idea and framework of higher-order adaptivity could be extended to study other adaptive higher-order systems, ranging from opinion formation to evolutionary dynamics \cite{liu2021modeling, friedkin2016network, alvarez2021evolutionary}. Besides, although the proposed higher-order adaptivity is representative, there still exist many other higher-order adaptive mechanisms, which is worthy of further explorations. For instance, threshold-like nonlinear relationship between structural evolution and the number of infected members, and flow of nodes among hyperedges of different scales \cite{burgio2025characteristic, iacopini2024temporal}.

{\it Acknowledgement} This work is supported by National Science and Technology Major Project (2022ZD0116800), Program of National Natural Science Foundation of China (12425114, 62141605, 12201026, 12301305, 62441617), the Fundamental Research Funds for the Central Universities, and Beijing Natural Science Foundation (Z230001)

\end{document}



\title{Supplementary Materials}





\maketitle
\tableofcontents
\newpage

\section{Data and Real Hypergraphs}
This work utilizes three real datasets, which respectively record contacts among individuals in different contexts, including a high school (Thiers13), a workplace (InVS15) and a primary school (LyonSchool) \cite{mastrandrea2015contact, genois2018can}. They can be downloaded from http://www.sociopatterns.org. Each dataset records face-to-face contacts with temporal resolution of 20s. Then, using similar methods to classical works \cite{iacopini2019simplicial}, we extract and augment real hypergraphs. First, we use a 5-minute temporal window to aggregate networks, and  treat maximal cliques as hyperedges. In this work, we limit simulations to occur on real hypergraphs only containing 2- and 3-hyperedges. Thus, hyperedges with large size are decomposed into 3-hyperedges. Meanwhile, according to their frequency, we just retain the top 20\% of hyperedges and derive a real-world hypergraph. Furthermore, to reduce stochastic fluctuations caused by finite size effect, we augment hypergraphs by 5 times through the algorithm of ref \cite{young2017construction}. It ensures that the augmented hypergraphs have the same structural characteristics with the original ones. In this work, we perform simulations on these augmented hypergraphs. Their statistical properties are as follows: (Thiers13) $N=1621$, $E_2=3570$, $E_3=4840$; (InVS15) $N=1081$, $E_2=5390$, $E_3=6020$; (LyonSchool) $N=1181$, $E_2=4790$, $E_3=3435$. Here, $N$, $E_2$ and $E_3$ respectively denote the number of nodes, 2-hyperedges and 3-hyperedges.

\section{Model Simulation}
We perform model simulations using a Gillespie algorithm. Suppose there are $n$ events, and each event $i$ occurs with rate $a_i$. According to Gillespie algorithm, only one event occurs at a time, and the probability that event $i$ occurs is $a_i/\sum_{j=1}^{j=n}a_j$.

Our model is composed of spreading process and hyperedge evolution, where each change in node state and hyperedge can be regarded as an event. We then compute the rate of these events. First, we consider spreading process. Within each hyperedge, the pathogen spreads to a susceptible node at rate $\theta(t) = \beta j^v$, where $j$ represents the number of infected nodes in the hyperedge. Each infected node recovers with rate $\mu$. Then, as for hyperedge evolution, each hyperedge containing at least 1 susceptible node breaks with rate $\pi(t) = r j^h$.

In each simulation, we set maximal iteration steps, which are sufficient to ensure that system reaches stationary state. Moreover, each point in figures is averaged over 30 runs.

\begin{figure}
\includegraphics[width=0.92\textwidth]{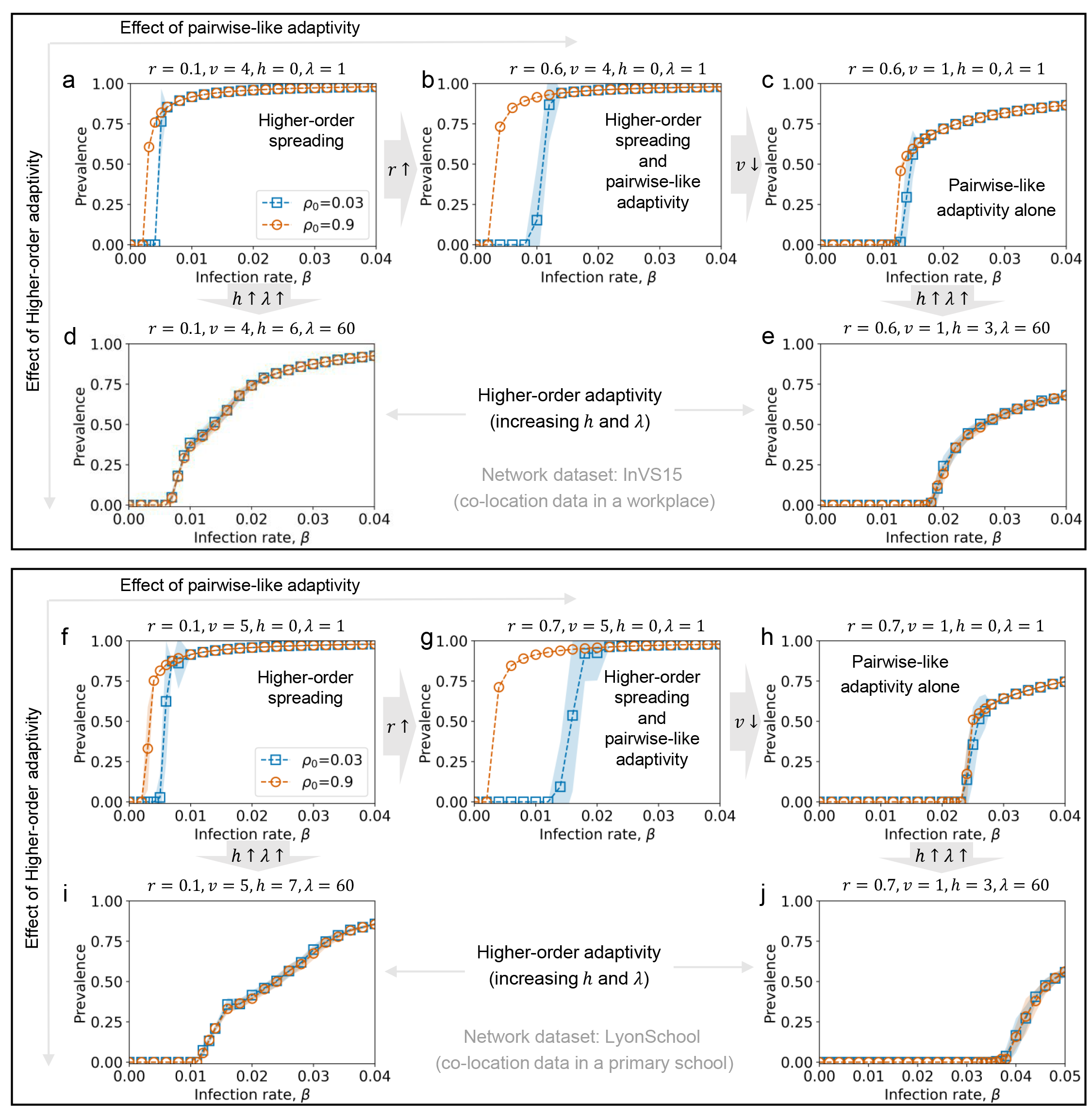}%
\caption{Results on two other real hypergraphs extracted from (a)--(e) InVS15 dataset and (f)--(j) LyonSchool dataset. (a)--(c) and (f)--(h) The impacts of pairwise-like adaptivity. Its strength is governed by parameter $r$. Results show that large $r$ can enhance or induce alone the bistable region and discontinuous phase transitions. (d)--(e) and (i)--(j) The impacts of higher-order adaptivity. It is depicted by parameters $h$ and $\lambda$. When increasing $h$ and $\lambda$, the discontinuous phase transitions and bistable region, regardless of whether they are induced by higher-order spreading mechanism or pairwise-like adaptivity, would disappear. Other parameter: $\mu=0.2$ \label{S1}}
\end{figure}

\section{Results on Two Other Real Hypergraphs}
To verify the robustness and generality of our findings, we preform simulations on two other real hypergraphs, which are respectively extracted from co-presence data in a workplace and a primary school \cite{genois2018can}. Firstly, we concentrate on the impacts of pairwise-like adaptivity. In the cases, hyperedge breaking rate has no dependence on the number of infected nodes, i.e., $h=0$, but rather is controlled solely by parameter $r$. Fig.\ref{S1}(a)--(c) and \ref{S1}(f)--(h) show simulation results on the two hypergraphs under this setting. Both of them demonstrate that pairwise-like adaptivity not only increases spreading thresholds but also enhance or induce bistable region and discontinuous phase transitions. The results are consistent with our findings in the main text.

Then, we consider the impacts of higher-order adaptivity. It is characterized by parameters $h$ and $\lambda$. Fig.\ref{S1}(d)--(e) present simulation results on a real hypergraph when the two parameters are set as higher values. Through comparisons between Fig.\ref{S1}(a) and \ref{S1}(d), as well as between Fig.\ref{S1}(c) and \ref{S1}(e), we find that higher-order adaptivity can lead to the disappearance of bistability and discontinuous phase transitions, regardless of whether the phenomena are induced by higher-order spreading mechanism and pairwise-like adaptivity. The results also hold when performing simulations on the other real hypergraph, as shown in Fig.\ref{S1}(i)--(j). These findings agree with results in the main text, which validates the robustness of higher-order adaptivity's impacts.

In summary, the impacts of pairwise-like adaptivity and higher-order adaptivity are robust and general across different underlying hypergraphs.

\section{Theoretical Analysis on 3-uniform Hypergraph}
\subsection{Dynamical Equations}
The main text provides general dynamical equations to depict co-evolutionary behaviors on any $m$-uniform hypergraph, as shown in Eqs.(5)--(6). In this part, we concentrate on the simple but representative higher-order situations, i.e., $m=3$, and perform theoretical analysis. In this case, the dynamical equations can be simplified as
\begin{equation}
\begin{aligned}
\frac{dN_i}{dt} &= -\mu N_i + (N-N_i)\Theta\\
\frac{dl_{3,1}}{dt} &=-2(\beta+\Theta)l_{3,1} - \mu  l_{3,1} + 3\Theta l_{3,0}+ 2\mu  l_{3,2} -r l_{3,1}+ 2\sum_{j=1}^{2}  rj^h l_{3,j} \Gamma (1-\Gamma)\\
\frac{dl_{3,2}}{dt} &=-(2^v \beta+\Theta) l_{3,2} - 2\mu  l_{3,2}+ 2(\beta +\Theta) l_{3,1}+ 3\mu l_{3,3} -2^h r l_{3,2}+ \sum_{j=1}^{2}  rj^h l_{3,j} (1-\Gamma)^2\\
\frac{dl_{3,3}}{dt} &=-3\mu  l_{3,3} + (2^v \beta + \Theta) l_{3,2}
\end{aligned}
\label{eqS1}
\end{equation}
where $N_i(t)$ denotes the number of infected nodes at time $t$ and $l_{3,k}(t)$ is the number of 3-hyperedges containing $k$ infected nodes. In addition, $\Theta$ represents the average rate that a $S$ node becomes infected, which satisfies
\begin{equation}
\Theta = \sum_{j=1}^{3} \frac{(3-j)l_{3,j}}{N-N_i} \beta j^v
\end{equation}
$\Gamma$ represents the possibility of selecting a susceptible node when forming a new hyperedge, which is written as
\begin{equation}
\Gamma = \frac{\lambda N_s}{\lambda N_s + (N-N_s)}
\label{gamma}
\end{equation}
where $N_s=N-N_i$ is the number of susceptible nodes.

Moreover, the total number of hyperedges, denoted by $E$, remains constant in our model. That is, the formula, i.e., $l_{3,0} = E-l_{3,1}-l_{3,2}-l_{3,3}$, holds at any time. Therefore, Eq.\eqref{eqS1} constitutes a self-contained 4-dimensional autonomous system.

\subsection{Theoretical analysis for the outbreak threshold}
In the autonomous system composed of Eq.\eqref{eqS1}, there is a trivial fixed point, which is $(N_i^*,l_{3,1}^*,l_{3,2}^*,l_{3,3}^*)=(0,0,0,0)$. It corresponds to the state that no individual is infected. In the initial stage of disease spreading, several individuals are infected, which can be regarded as small perturbations around the fixed point. If the fixed point is stable, the perturbations will converge to zero, which means that the system returns to the state with no infected individuals. Otherwise, the disease will outbreak. Therefore, the stability condition of this fixed point corresponds to the outbreak threshold of epidemics.

Here, we analyze the stability condition by using linearization method and Routh-Hurwitz stability criterion. Firstly, we linearize the system around the fixed point (0,0,0,0) and obtain the Jacobian Matrix, which reads as
\begin{equation}
J_{(0,0,0,0)} = \left(\begin{array}{cccc}
-\mu & 2\beta & 2^v \beta & 0\\
0 & -2\beta-\mu+ 2\langle d\rangle\beta-r & 2^v \langle d\rangle \beta + 2\mu & 0\\
0 & 2\beta & -2^v \beta - 2\mu-2^h r & 3\mu\\
0 & 0 & 2^v \beta & -3\mu\\
\end{array}\right)
\label{eqS4}
\end{equation}
where $\langle d\rangle = 3E/N$ represents the average hyperdegree. As we all know, the fixed point is stable if and only if all eigenvalues are negative or have negative real parts. Note that $-\mu$ is a negative eigenvalue. Thus, we only need to analyze the remaining three eigenvalues, which are determined by the last three rows of the matrix \eqref{eqS4}, that is,
\begin{equation}
\left(\begin{array}{ccc}
a_{11} & a_{12} & 0 \\
a_{21} & a_{22} & a_{23} \\
0 & a_{32} & a_{33}
\end{array}\right) = \left(\begin{array}{ccc}
-2\beta-\mu+ 2\langle d\rangle\beta-r & 2^v \langle d\rangle \beta + 2\mu & 0\\
2\beta & -2^v \beta - 2\mu-2^h r & 3\mu\\
0 & 2^v \beta & -3\mu\\
\end{array}\right)
\label{eqS5}
\end{equation}

Then, the characteristic equation of Matrix \eqref{eqS5} is as follows:
\begin{equation}
\begin{aligned}
0 =&-s^3+(a_{11}+a_{22}+a_{33})s^2\\
&-(a_{11}a_{22}+a_{11}a_{33}+a_{22}a_{33}-a_{12}a_{21}-a_{23}a_{32})s\\
&+a_{11}a_{22}a_{33}-a_{12}a_{21}a_{33}-a_{11}a_{23}a_{32}
\end{aligned}
\label{eqS6}
\end{equation}

To judge whether all solutions have negative real parts, we then utilize the Routh-Hurwitz stability criterion. Specifically, we first calculate the Routh table, which is written as
\begin{equation}
\left(\begin{array}{cc}
-1 & y \\
a_{11}+a_{22}+a_{33} & a_{11}a_{22}a_{33}-a_{12}a_{21}a_{33}-a_{11}a_{23}a_{32} \\
\frac{-a_{11}a_{22}a_{33}+a_{12}a_{21}a_{33}+a_{11}a_{23}a_{32}}{a_{11}+a_{22}+a_{33}}-y & 0\\
-a_{11}a_{22}a_{33}+a_{12}a_{21}a_{33}+a_{11}a_{23}a_{32} & 0\\
\end{array}\right)
\label{eqS7}
\end{equation}
where $y=-a_{11}a_{22}-a_{11}a_{33}-a_{22}a_{33}+a_{12}a_{21}+a_{23}a_{32}$. Routh-Hurwitz stability criterion suggests that all solutions have negative real parts if and only if the values at the first column of the Routh table are all negative. Thus, the stability condition can be computed by these inequalities
\begin{equation}
\begin{aligned}
a_{11}+a_{22}+a_{33}<0\\
\frac{-a_{11}a_{22}a_{33}+a_{12}a_{21}a_{33}+a_{11}a_{23}a_{32}}{a_{11}+a_{22}+a_{33}}-y <0 \\
-a_{11}a_{22}a_{33}+a_{12}a_{21}a_{33}+a_{11}a_{23}a_{32}<0
\end{aligned}
\label{eqS8}
\end{equation}

We then substitute Eq.\eqref{eqS5} into Eq.\eqref{eqS8}. Through calculation and comparison with numerical solutions of Eq.\eqref{eqS1}, we find that the last formula of Eq.\eqref{eqS8} dominates the outbreak threshold, which can be simplified as
\begin{equation}
-2^{v+1}\langle d\rangle\beta^2 + [2^{h+1}r (1-\langle d\rangle)-4\mu\langle d\rangle]\beta+(\mu+r)(2\mu+2^h r)>0
\label{eqS9}
\end{equation}

We notice that the left expression is a quadratic function with respect to $\beta$ and that the coefficient of the quadratic term $-2^{v+1}\langle d\rangle<0$. In addition, according to Vieta's formulas, the product of two roots $\beta^*_1\beta^*_2 = -(\mu+r)(2\mu+2^h r)/2^{v+1}\langle d\rangle<0$. It indicates that the two roots are one positive and one negative. In this way, we can obtain the figure of the quadratic function, and further conclude that the positive root is the outbreak threshold, which is given by
\begin{equation}
\beta_c = \frac{\Omega + \sqrt{\Omega^2+2^{v+3}\langle d\rangle(\mu+r)(2\mu+2^h r)}}{2^{v+2}\langle d\rangle}
\label{eqS10}
\end{equation}
where
\begin{equation}
\Omega = 2^{h+1}r(1-\langle d\rangle)-4\mu \langle d\rangle
\end{equation}
\section{Supplement study about higher-order adaptivity}
\subsection{Impacts on phase transitions}
\begin{figure}
\includegraphics[width=0.9\textwidth]{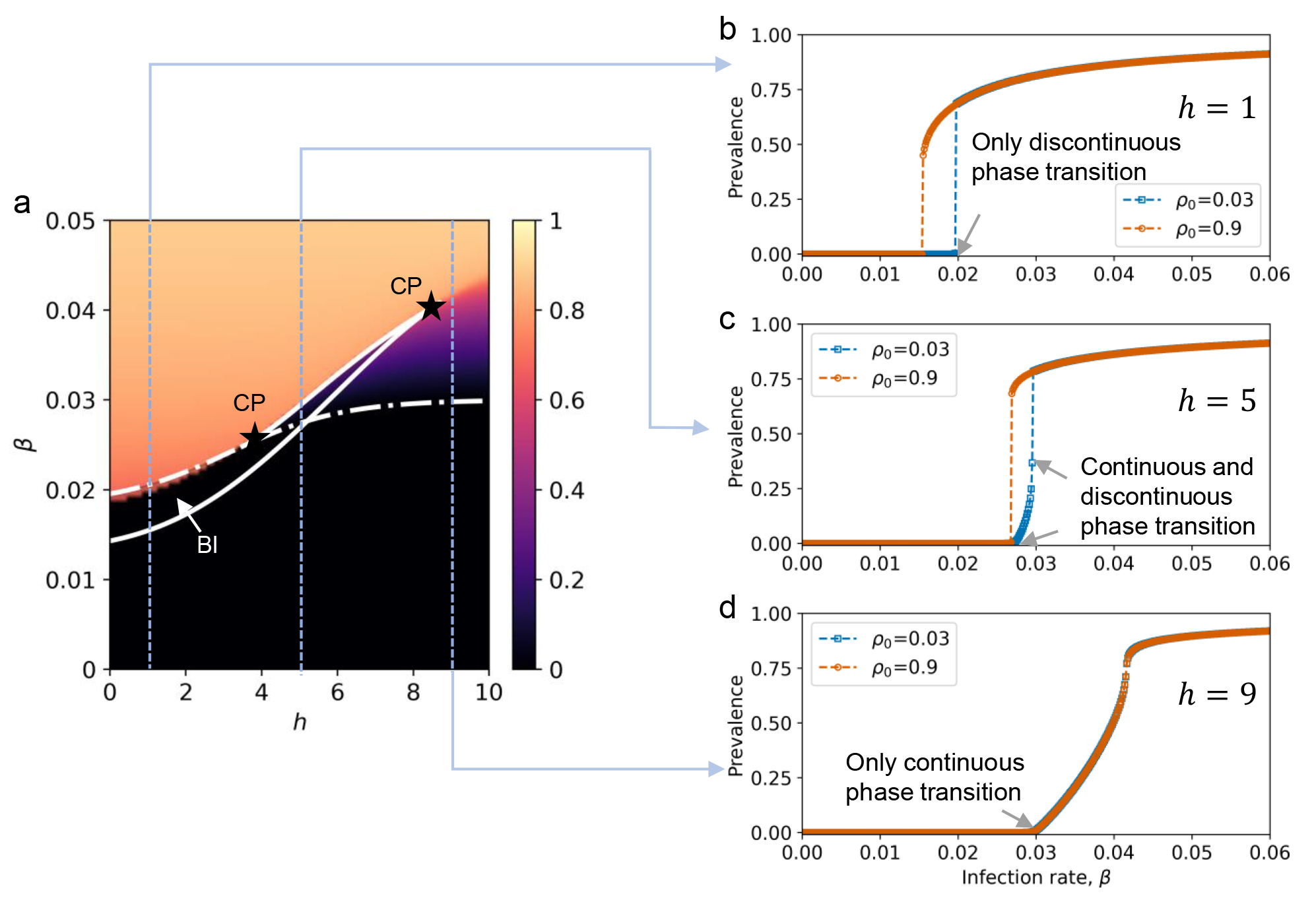}%
\caption{Effects of parameter $h$ on phase transitions. (a) Phase diagram of epidemic prevalence at small $\rho_0$. The dashed line is calculated by Eq.\eqref{eqS10} and predicts the outbreak threshold well. The solid line represents persistence threshold or drastic-change threshold, numerically solved by Eq.\eqref{eqS1}. Moreover, there are two cusp points (CP), which are derived by ManCont \cite{dhooge2008new}. (b)--(d) Prevalence curves against $\beta$. We select three $h$ values before and after CPs: (b) $h=1$, (c) $h=5$ and (d) $h=9$.  As $h$ grows, results intuitively present the changes of the number and the type of phase transitions. Other parameters: $v=3, r=0.1, \mu=0.2, \lambda=1$. \label{S2}}
\end{figure}

This session provides a detailed presentation about how $h$ influences the number and types of phase transitions, as the supplement of Fig.3. Fig.\ref{S2}(a) presents phase diagrams of epidemic prevalence when the initial density of infected individuals ($\rho_0$) is small, which is a complementary figure of Fig.3(h). The dashed line is calculated by Eq.\eqref{eqS10}, which could well predict the outbreak threshold. Moreover, Fig.\ref{S2}(a) illustrates again the pivotal role of $h$ in increasing spreading thresholds, shrinking bistable region and triggering the emergence of cusp points (CP) where the number and type of phase transitions alters. For intuitive presentation, we select three representative values of $h$ before and after CPs, and plot prevalence curves against $\beta$ in Fig.\ref{S2}(b)--(d). We find that as $h$ grows, phase transition changes from a discontinuous one ($h=1$) to the co-existence of continuous and discontinuous ($h=5$), and then to a continuous one ($h=9$).

\subsection{Impacts when excluding the influence of higher-order spreading}
\begin{figure}
\includegraphics[width=0.7\textwidth]{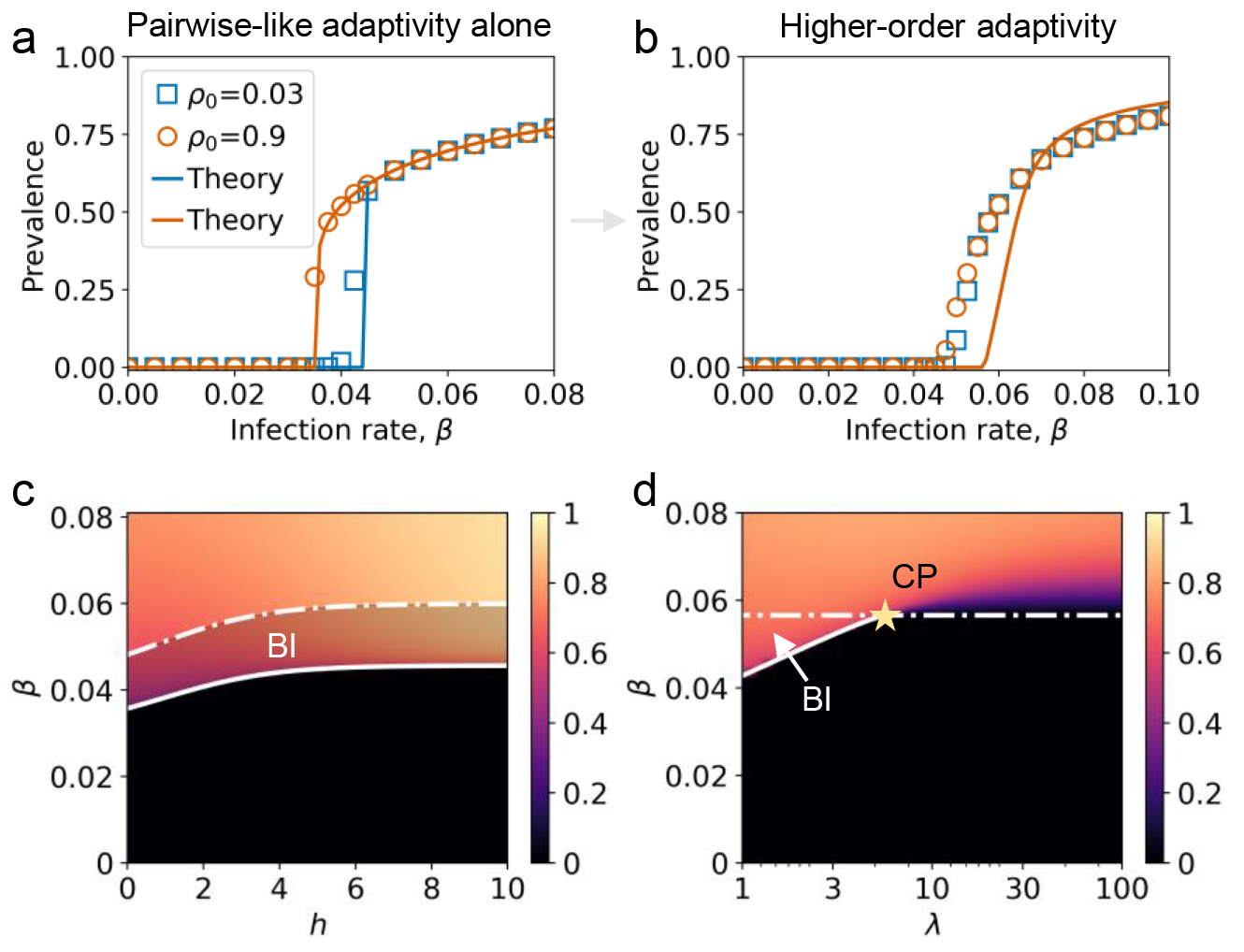}%
\caption{Impacts of higher-order adaptivity when excluding the influence of higher-order spreading. (a)--(b) Shown is epidemic prevalence under (a) $r=0.4, h=0, \lambda=1$ and (b) $r=0.4, h=3, \lambda=60$. (c)--(d) Phase diagrams under different settings: (c) $r=0.4, \lambda=1$; (d) $r=0.4, h=3$. The dashed and solid line respectively represent outbreak threshold and persistence threshold. The area they enclose is bistable region (BI) and the star is cusp point (CP). We find that when bistable region is induced by pairwise-like adaptivity, its shrinkage caused by higher-order adaptivity is mainly determined by $\lambda$. Other parameters: $N=3000, E=6000, \mu=0.2$. \label{S3}}
\end{figure}

Here, we explore the impacts of higher-order adaptivity when higher-order spreading mechanism is removed ($v=1$). From Fig.\ref{S3}(a) to Fig.\ref{S3}(b), we show that higher-order adaptivity could still eliminate the bistability and discontinuous transitions induced by pairwise-like adaptivity ($r=0.4$). Furthermore, Fig.\ref{S3}(c) and Fig.\ref{S3}(d) present that the bistable region remains almost unchanged as $h$ grows while vanishes as the fitness parameter $\lambda$ increases. This indicates that the inhibitory effect of higher-order adaptivity is largely determined by $\lambda$ in the case, which is different from the scenarios where higher-order spreading plays a role (see Fig.3 in the manuscript).

